# Ferromagnetic Two-dimensional Electron Gases with Magnetic Doping and Proximity Effects

Zixin Fan[1], Jiale Chen[2,3], Qiangtao Sui[1], Haoming Ling[1], Zihao Wang[1], Lingyuan Kong[1], Dingyi Li[1], Fang Yang[1], Run Zhao[4,5], Hanghui Chen[3,6], Pan Chen[1*], Yan Liang[1*] and Jiandi Zhang[1,7*]

[1]*Beijing National Laboratory for Condensed Matter Physics, Institute of Physics, Chinese Academy of Sciences, Beijing 100190, China*

[2]*Department of Electronic Science, East China Normal University, Shanghai 200241, China*

[3]*NYU-ECNU Institute of Physics, NYU Shanghai, Shanghai, 200124 China*

[4]*School of Physical Science and Technology, Suzhou University of Science and Technology, 215009, Suzhou, China*

[5]*Key Laboratory of Intelligent Optoelectronic Devices and Chips of Jiangsu Higher Education Institutions, School of Physical Science and Technology, Suzhou University of Science and Technology, Suzhou, 215009, China*

[6]*Department of Physics, New York University, New York, NY, 10012 USA*

[7]*College of Physics, University of Chinese Academy of Sciences, Beijing, 100190, China*

**Abstract**

The advent of magnetic two-dimensional electron gases (2DEGs) at oxide interfaces has provided new opportunities in the field of spintronics. The enhancement of magnetism in 2DEGs at oxide interfaces continues to be a significant challenge, as exemplified by the relatively weak magnetism observed in the classical $LaAlO_3/SrTiO_3$ interface. Here, we present ferromagnetic (FM) 2DEGs at the interface fabricated between the FM insulator $EuTiO_3$ (ETO) and the strong spin-orbit coupled (SOC) perovskite insulator $KTaO_3$ (KTO). With the combined effects of magnetic atom doping and magnetic proximity from ETO films, the coercive field of 2DEGs can be significantly enhanced. Magnetoresistance (MR) curve with a high coercive field of 1000 Oe has been observed, in conjunction with a temperature-dependent unambiguous hysteresis loop in the anomalous Hall effect (AHE). Furthermore, within the 2DEGs, we have identified a synergetic interplay between magnetic scattering and the weak antilocalization (WAL) effect on transport. This study provides fresh insights into the formation of FM 2DEGs at ETO/KTO interfaces, and introduce an innovative pathway for creating high-performance magnetic 2DEGs at oxide interfaces.

**Introduction**

Since the discovery of interfacial two-dimensional electron gases (2DEGs) between insulating oxides[1], significant attention has been drawn to this field. Various phenomena associated with 2DEGs, including superconductivity[2-4], quantum oscillation[5, 6], Rashba spin-orbit coupling[7, 8] (SOC), ferromagnetism[9, 10], anomalous Hall effect[11-13] (AHE), and anisotropic hysteresis, have spurred advancement in oxide electronics. Notably, the investigation of magnetic properties in 2DEGs is a central area of research. The discovery of magnetic 2DEGs at the $LaAlO_3/SrTiO_3$ (LAO/STO) interface in 2007 represented a significant advancement in the field of oxide electronics[14], despite the transition temperature ($T_C$) been a mere 300 mK. Since then, enhancing magnetism in 2DEGs has become a primary research focus. Various methods, including magnetic impurity doping[15], the insertion of thin ferromagnetic (FM) layers[11, 16], and the utilization of the magnetic proximity effect[9, 13], have been explored to achieve magnetic enhancement. Remarkably, a magnetic 2DEG with a $T_C$ of 70 K was observed at the interface of EuO and $KTaO_3$ (KTO), facilitated by the magnetic proximity effect. However, the coercive field, as derived from the MR curve, within the 2DEG is approximately mere 200 Oe. This makes it challenging to observe a hysteresis loop in the AHE curve. In addition to the EuO/KTO heterostructure, existing literature consistently reports extremely small coercive fields in various systems such as $EuTiO_3$ (ETO)/STO[17], EuO/STO[13], and $LaTiO_3$/STO[18]. In these systems, magnetism predominantly stems from a singular source, whether it be magnetic atom doping or the effect of magnetic proximity, leading to the relatively low coercive field observed in the 2DEG and a hardly discernible hysteresis loop in AHE curve.

Perovskite-type KTO substrates are prone to losing the A-site element potassium (K) at high temperatures. This vulnerability offers a viable method for introducing magnetic atoms to the surface of KTO substrates. Given the doping of europium (Eu) atoms in KTO substrates, together with the FM insulating characteristic of ETO films, the 2DEG at the interface of ETO/KTO heterostructure is likely to exhibit strong magnetism [Fig. 1]. In this study, single-crystal ETO films were grown on KTO (001) substrates using the pulsed laser deposition (PLD) method to generate FM 2DEGs at the interfaces between them. The doping of magnetic Eu atoms, in conjunction with the proximity effect arising from the FM ETO film, plays a pivotal role in shaping the FM nature of 2DEGs. The magnetoresistance (MR) curves exhibiting a high coercive field of 1000 Oe has been demonstrated, in accompanied with an unusual temperature-dependent hysteresis loop evident in the AHE curves. The AHE curves shows a hump-shaped structure which can be attributed to the presence of multiple magnetic layers/domains, which correspond to the doping layers of varying Eu doping concentration within the substrate's surface layer. In 2DEGs, the scattering of electrons in Ta $5d$ orbitals by the Eu impurity moments manifests in signatures of Kondo effect observed in low-temperature transport. This has been confirmed through density functional theory (DFT) calculations. Besides, the robust SOC in KTO substrates has led to the weak antilocalization (WAL) effect in the 2DEG, leading to a synergetic interplay between magnetic scattering and the WAL effect at low temperatures.

**Result and discussion**

ETO thin films were grown on (001)-oriented KTO single crystals using a PLD system with a base pressure below $5\times10^{-10}$ Torr, see the Methods section for details. Bulk ETO phase crystalizes in a cubic structure with a lattice constant of 3.904 Å, KTO is also cubic with a lattice constant of 3.989 Å, thereby ensuring high-quality epitaxy [Fig. 1(a)]. X-ray diffraction [XRD, Fig. S1(a)] and φ-scanning [Fig. S1(b)] results demonstrate the formation of single-crystal ETO film with four-fold symmetry on the KTO surface, characterized by an ultra-smooth surface with a root-mean-square (RMS) roughness of merely 0.25 nm. Through careful analysis and fitting of the small-angle X-ray reflectance (XRR) profile [Fig. S1(c)], we have determined that the thickness of the ETO film is approximately 10.7 nm. Furthermore, our investigations reveal the existence of an interlayer between the ETO and KTO layers, with an approximate thickness of 1.5 nm. The High Angle Annular Dark Field (HAADF) image [Fig. 1(b), Fig. S3 (a)], derived from atomic-resolution Scanning Transmission Electron

Microscopy (STEM), validates the occurrence of interlayer at the interface of ETO and KTO, indicated by two red dashed lines. The atomic line profile of atoms at site A is represented by a yellow peak array. Fig. S2 (b) left illustrates the concentration of Eu and K atoms, the values of which are derived from EELS spectra. Fig. S2 (b) right provides a magnification of the interface depicted in Fig. S2 (a). The results from both the line profile and EELS spectra suggest the presence of an interlayer, with an approximate thickness of 1.5 nm, between ETO and KTO. Furthermore, as shown in Fig. S3, the interlayer contains Eu, K, Ta, and Ti atoms.

The ETO thin films obtained here exhibit the standard magnetic properties of the strained ETO phase[19]. Additionally, the interlayer region, as previously mentioned, exhibits magnetic properties distinct from those of the ETO films. The magnetism of the ETO/KTO heterostructure was measured using a Magnetic Property Measurement System (MPMS), with both external magnetic fields perpendicular and parallel to the sample surfaces. As shown in Fig. 2(a), the temperature-dependent magnetization starts to rearrange under an applied magnetic field of 0.1 T parallel (perpendicular) to the film surface, and undergoes a magnetic transition around 7 K, which closely resembles those of previously reported epitaxial ETO films[20-22]. At 2 K, the film shows pronounced magnetic anisotropy, with the easy axis of magnetization residing within the film plane, with saturation magnetization of 6.9 $\mu_B$/Eu [Fig. S1(d)]. The coercive field of ETO film with out-of-plane magnetic field is about 400 Oe. It is noteworthy that the in-plane *M-H* hysteresis exhibits more than one loop shape [Fig. 2(b)], suggesting the presence of multiple magnetic domains within the ETO/KTO heterostructure[23,24]. As ETO film is a single crystal ferromagnetic insulator, other magnetic domains may come from the 2DEGs at the interface of ETO/KTO [Fig. S4(a)]. The superposition of different hysteresis loops results in the double loops hysteresis curves. The double loops become increasingly evident as reducing film thickness [Fig. S4(b)], which illustrates that the contributions from 2DEGs to the total magnetism are progressively becoming more significant. This observation further substantiates the occurrence of an interlayer at the ETO/KTO interface, with its magnetic properties differing from those of the ETO film[20, 25-27].

Figure 2(c) shows the temperature-dependent sheet resistance (*R-T*) curve of the ETO/KTO interface. Two features can be identified from the *R-T* curve: 1) the 2DEG exhibits metallic characteristics when cooled from 300 K to 35 K; 2) a pronounced resistance upturn appears below 35 K. By employing the model [$R(T) = R_0 - q\ln T + R_K(T)$] that integrates the impact of $\ln T$ and the Kondo effect [$R_K(T)$], we are able to precisely fit the upward trend at low temperatures in the *R-T* curve (the red solid curve). $\ln T$, reflecting the unsaturated resistance increase, typically arises from WL and electron-electron interactions. Given the strong SOC in KTO and the location of the 2DEGs on the KTO surface, we can infer that WAL inevitably exists within the 2DEGs [which is evident in the MR data rather than R(*T*)]. Since WL and WAL are mutually exclusive, we can rule out the presence of WL in the system. Therefore, the $\ln T$ dependence is primarily due to electron-electron interactions. The gray solid curve in Fig. 2(c) is the fitting to the data by $\ln T$ solely, which shows significant deviation of the experimental data from the fitting in low-*T* regime. The term $R_K(T)$ is account for the contribution of the Kondo effect, which is mathematically expressed as follows[28,29]:

$$R_K(T) = R_K(T=0)\left[\left(\frac{T}{T_K}\right)^2 \left(2^{\frac{1}{s}} - 1\right) + 1\right]^{-s}$$

The function $R_K(T)$ is universally dependent on the *T* in units of the Kondo temperature $T_K$, defined as the temperature at which the Kondo effect contributes half of its zero-temperature resistance value. The parameter *s*, which is dependent on the impurity spin (*S*), is conventionally set to *s* = 0.225 for *S* = ½ . The fitting parameters are listed in Table S1. The observation that the *R-T* curve aligns accurately following the incorporation of the Kondo effect and $\ln T$ supports our claim regarding the presence of the Kondo effect in the ETO/KTO system.

The Kondo effect in the 2DEGs between oxides can be induced by two primary factors: magnetic impurities and a substantial quantity of oxygen vacancies, which can potentially generate magnetic centers [28] . To elucidate the origin of Kondo effect in the ETO/KTO system, we conducted a comparative experiment of STO/KTO. The STO/KTO heterostructure was developed under identical conditions to those of the ETO/KTO (Fig. S5). Notably, the *R-T* curve of STO/KTO aligns well with the $\ln T$ trend at low temperatures [as depicted by the green dotted curve and the gray fitting line in Fig. 2(c)], suggesting the absence of a Kondo effect in the STO/KTO system. Given that the KTO substrates are prepared under consistent conditions in ETO/KTO and STO/KTO, we can reasonably eliminate the influence of oxygen vacancies. Furthermore, as the STO film is nonmagnetic, it can be inferred that magnetic atoms (Eu) have played a pivotal role in the Kondo effect observed within the ETO/KTO system.

The Hall resistance of the 2DEG at the ETO/KTO interface is measured by applying a magnetic field perpendicular to the film surface. The Hall resistance presents a robust linear variation under the 9 T external magnetic field (Fig. S6), suggesting the presence of a single type of carrier at the interface. The temperature-dependent carrier density and mobility of the 2DEG are illustrated in Fig. 2(d). As the test temperature decreases from 300 K to 2 K, there is a consistent increase in carrier mobility and a corresponding decrease in carrier concentration.

A remarkable observation is the hysteretic magneto transport behaviors in the 2DEG at the ETO/KTO interface, suggesting that the 2DEG possesses FM properties. Fig. 3(a) illustrates the MR of the 2DEG collected at 2 K to 10 K while cycling magnetic field along the +2000 Oe to –2000 Oe and to +2000 Oe. MR, which denotes the alteration in resistance in response to an applied magnetic field, is quantified by the formula MR = [$R$(B) - $R$(0)]/$R$(0). Here, $R$(B) represents the resistance value under a magnetic field B, while $R$(0) signifies the resistance value at zero magnetic field. At 2 K, two neighboring MR minima located at ±1000 Oe are observed, forming a butterfly-shaped MR-$H$ curve. The MR minima field is defined as the coercive field of the FM 2DEG. The value of this field in the 2DEG is nearly the highest among those reported in literatures[9, 13, 17]. The appearance of hysteretic MR strongly suggests the establishment of FM 2DEG at the ETO/KTO interface. Increase in temperature causes a rapid decrease in MR, which becomes indistinct at 5 K. This suggests that FM transition temperature of the 2DEG is approximately 5 K. Given that the $T_C$ of epitaxial ETO film is around 7 K, we infer that the magnetic nature of the 2DEG is likely induced by the proximity effect originating from the ETO film. As the magnetic proximity effect is considered to be a weak interaction. The magnetism transferred via the proximity effect is significantly mitigated, leading to a decrease in $T_C$ of the 2DEG. However, the coercive field of ETO film with out-of-plane magnetic field is about 400 Oe [Fig. 2(b)], which is much smaller than that of the 2DEG (1000 Oe). The findings suggest that the proximity effect is a contributing factor to the ferromagnetism in 2DEGs, however, it is not the sole origin. In fact, the enhancement of ferromagnetism in the 2DEG can be attributed to both the proximity effect and the magnetic atoms doping effect.

More particularly, we have observed temperature-dependent AHE that exhibits unambiguous hysteresis loops in the 2DEG. In Hall effect measurements, we employed a perpendicular cyclic external magnetic field ranging from -5000 Oe to 5000 Oe. After the subtraction of the ordinary Hall effect component, a distinct hump-like AHE emerged [Fig. 3(b)], the coercive field in AHE is proximate to that derived from the MR curves. As the measuring temperature increases, the total intensity of AHE and coercive field continuously decrease, while the hump initially increases and then decreases (Fig. S7). This observed behavior markedly contrasts to the AHE seen in other magnetic 2DEG systems[9], which barely exhibits hysteresis loops or hump structure. We propose that these hump-like Hall peaks in the 2DEG are more likely associated with the presence of magnetic multilayer/domain structures[30-33], rather than the skyrmion-induced topological Hall effect suggested in various heterostructures[34, 35]. These multiple magnetic domain effects, such as the simplest two domain scenario with opposite signs of AHE as schematically depicted in Fig. 3(c), can qualitatively reproduce the experimental AHE. An analysis of the temperature-dependent curve of hump, reveals that the hump attains its apex at 3 K [Fig. 3(d)]. This behavior distinctly differs from the linear temperature dependence of hump intensity predicted by the dual AHE model. As previously noted, there is a gradual decrease in the concentration of Eu from the first to the third layer within the interlayer, thereby creating multiple magnetic domains in the 2DEGs [as schematically shown in Fig. S8]. This variation leads to the emergence of diverse AHE curves, which, when superimposed, result in a hump-shaped AHE curve with non-linear temperature dependence[36, 37].

In the 2DEG, with ferromagnetism and SOC, we have also observed a synergetic interplay between magnetic scattering and the WAL effect. High-field MR, ranging from -9 T to 9 T, exhibits a significantly intricate behavior [Fig. 4(a)]. Such complex MR behavior suggests that the 2DEG is affected by more than a single effect. The complex MR can be described by the following formula:

$$MR = \Delta R_{CO} + \Delta R_{EEI} + \Delta R_{MAG} + \Delta R_{WAL} \quad (1)$$

$\Delta R_{co}$ signifies the classical orbital contribution[38] (CO) to MR, $\Delta R_{EEI}$ represents the electron-electron interactions[39] (EEI). Both the $\Delta R_{co}$ and the $\Delta R_{EEI}$ are proportional to $B^2$ and are amalgamated into the term $AB^2$ of the equation. $\Delta R_{Mag}$ corresponds to the magnetic atoms scattering and magnetic proximity contribution, which exerts a negative influence on MR[40], describes by the term $-\frac{CB^2}{D+EB^2}$. $\Delta R_{WAL}$ stands for WAL, which is caused by strong SOC, can be expressed by the Iordanski, Lyanda-Geller, and Pikus (ILP) theory[41, 42].

$$\frac{\Delta \sigma(B)}{\sigma_0} = -\frac{1}{2}\psi\left(\frac{1}{2} + \frac{B_\varphi}{B}\right) + \psi\left(\frac{1}{2} + \frac{B_\varphi + B_{so}}{B}\right) + \frac{1}{2}\psi\left(\frac{1}{2} + \frac{B_\varphi + 2B_{so}}{B}\right)$$
$$+ \frac{1}{2}ln\left(\frac{B_\varphi}{B}\right) - ln\left(\frac{B_\varphi + B_{so}}{B}\right) - \frac{1}{2}ln\left(\frac{B_\varphi + 2B_{so}}{B}\right) \quad (2)$$

The formula involves the digamma function $\psi$, the quantum of conductivity denoted as $\sigma_0 = e^2/\pi h$, and the effective fields $B_\varphi$ and $B_{so}$ related to the inelastic and spin-orbit relaxation lengths, respectively. Given that the contributions from EEI and CO effects are invariably proportional to $B^2$, the respective weights of magnetic scattering and WAL contributions are incorporated to elucidate their relative proportions within the MR at varying temperatures. Consequently, equation (2) is adapted into the subsequent formulation:

$$MR = AB^2 + \sigma_0\left\{M\left[-\frac{CB^2}{D+EB^2}\right] + N\left[-\frac{1}{2}\psi\left(\frac{1}{2} + \frac{B_\varphi}{B}\right) + \psi\left(\frac{1}{2} + \frac{B_\varphi+B_{so}}{B}\right) + \frac{1}{2}\psi\left(\frac{1}{2} + \frac{B_\varphi+2B_{so}}{B}\right) + \frac{1}{2}ln\left(\frac{B_\varphi}{B}\right) - ln\left(\frac{B_\varphi+B_{so}}{B}\right) - \frac{1}{2}ln\left(\frac{B_\varphi+2B_{so}}{B}\right)\right]\right\}^{-1} \quad (3)$$

where M and N are the coefficients that weight the contributions from magnetic scattering and WAL. Taking the MR at 2 K as an example, the fitting results derived solely from ILP theory are incongruent with the empirical data. However, employing the weighted equation (3) yields a fitting curve that aligns well with the observed data [Fig. S9(a)]. The fitting result suggests that both WAL and magnetic scattering significantly contribute to the MR at low temperature. Upon incrementally elevating the temperature to 10 K and 20 K, the positive magnetoconductance becomes negligible, indicating the disappearance of magnetic scattering contributions [Fig. 4(b)].

To evaluate WAL and magnetic contributions at various temperatures, coefficients N and M from equation (5) were used to define the ratio r=N/M [Fig. 4(d)]. Analysis of the effective inelastic field ($B_\varphi$) and inelastic scattering length ($l_\varphi$) revealed that WAL intensity decreases with rising temperature [Fig. S9(b)]. Notably, under a constant temperature, an increase in magnetic field weakens the WAL effect and enhances magnetic scattering until saturation. At 2 K, the value of r = 0.98 indicates that the WAL and magnetic scattering show similar contributions to MR [Fig. 4(d)]. As shown in Fig. 4(c), within a range of ±0.88 T at 2 K, the contribution from WAL exceeds that of magnetic scattering, thereby maximizing the MR. Outside of this range, magnetism prevails, leading to a decrease in MR. At 5 K, the MR peak narrows to ±0.2 T, where there is a balance between WAL and magnetic scattering from ±0.2 T to ±0.45 T. Beyond ±0.45 T, magnetic scattering dominates, causing a decrease in MR until it reaches a minimum at ±1.12 T. Here, r = 0.73 suggests a slightly higher magnetic contribution. At 10 K and 20 K, the r values of 33.16 and 355.77 indicate magnetic disappearance and WAL dominance respectively, resulting in the transformation of the MR-H curve into a consistently positive magnetoresistance feature [Fig. 4(d)]. When the temperature either equals or exceeds 50 K [Fig. 4(a)], WAL vanishes as well, leading to the transformation of the MR-H curve into a fully positive magnetoresistance characteristic.

To understand the effects of the Eu atoms within the 2DEGs, we perform first-principles calculations (the computational details can be found in the Method). The cross-sectional Electron Energy Loss Spectroscopy (EELS) results, acquired from the interface of ETO/KTO for K, Eu, Ta, and Ti, have confirmed the postulation that the doping of Eu atoms at the interface is gradient (Fig. S10). As shown in Fig. S3, the Eu concentration is approximately 75%, 50%, and 25% from the first, second and third layer within the interlayer. In our DFT calculations, for computational efficiency, we selected a 25% concentration. Additionally, given the limited functionality of Ti, it will not be included in our calculations. We simulate interlayer by using a prototypical supercell of Eu$_{0.25}$K$_{0.75}$TaO$_3$. We use $U = 4.5$ eV on Eu-$f$ orbitals because it reproduces the insulating $G$-type antiferromagnetic (AFM) ground state of bulk ETO and is also close to the AFM-to-FM phase boundary [Fig. S10 and Fig. S11]. In the supercell of Eu$_{0.25}$K$_{0.75}$TaO$_3$, there are three configurations labeled as (I), (II) and (III) [Fig. 5(a)]. In the configuration (I), the two Eu atoms are adjacent with a separation of about one lattice constant $a$; in the configuration (II), the two Eu atoms are along the face-diagonal line with the separation of about $\sqrt{2}a$; in the configuration (III), the two Eu atoms are along the body-diagonal line with the separation of about $\sqrt{3}a$. For each configuration, we align the Eu-$f$ moments either ferromagnetically or anti-ferromagnetically. Our calculations find that configuration (I) has the lowest total energy among the three configurations, implying that there is a tendency for Eu clustering in Eu$_x$K$_{1-x}$TaO$_3$. Furthermore, as illustrated in Fig. 5(b) and Fig. S11, for each configuration, the FM and AFM states always have very close energies (the energy difference is smaller than 1 meV per Eu). This suggests that the magnetic exchange interaction between Eu-$f$ moments is minimal, making it improbable for Eu-$f$ moments to establish a long-range order. Even if an order is formed, the transition temperature would be exceedingly low. Figure 5(c) and 5(d) shows the band structure and density of states of Eu$_{0.25}$K$_{0.75}$TaO$_3$ in the FM and AFM states, respectively. Since the valence of Eu is 2+ and the valence of K is 1+, the substitution of Eu with K in KTO acts as electron doping. Therefore, the Fermi level moves above the conduction band and itinerant electrons fill the Ta-$d$ states. The Eu-$f$ states do not cross the Fermi level.

The above calculation results suggest that Eu-doped KTO exhibits metallic properties with the Eu-$f$ moments coupling to itinerant electrons in the Ta-$d$ bands, which leads to the Kondo effect. This explains the upturn of the resistivity at low temperatures in the transport measurements. However, the verification of the proximity effect of FM ETO is currently computationally challenging due to the immense computational load. While taking into account the consistency of the coercive fields between the thin ETO film and 2DEG, in conjunction with reported literatures, it can be inferred that ETO exerts a proximity effect on the 2DEG. By synthesizing both computational and experimental observations on the impacts of magnetic atom doping, it can be concluded that the combined effects of magnetic atoms doping and the proximity effect result in an enhancement of ferromagnetism in the 2DEG within the ETO/KTO system.

**Conclusion**

We have successfully fabricated an FM 2DGE with enhanced coercive field at the interface of an insulating FM ETO thin film and a KTO substrate, a $5d$ transition metal oxide renowned for its robust SOC. By intentionally creating K vacancies on the KTO substrate, a magnetic conducting layer populated by Eu atoms is formed during film deposition. This is evidenced by the STEM results, the observation of Kondo effect, and DFT calculations. The doping of magnetic Eu atoms, in conjunction with the proximity effect arising from the FM ETO film, plays a pivotal role in shaping the FM nature of 2DEGs. Consequently, a high coercive field of 1000 Oe has been observed in the 2DEG, which is further evidenced by a distinct hysteresis loop in the AHE. Concurrently, we have observed an interplay between strong magnetism and WAL in the 2DEG, resulting in intricate high-field MR behavior. Our research provides a universal approach to magnetic enhancement, integrating the doping of magnetic atoms with the proximity effect, offering unique opportunities for future applications in two-dimensional spintronics.

**Methods**

The ETO thin films were deposited on a single-polished KTO (001) single crystal with size of 2.5 mm × 5 mm × 0.5 mm utilizing PLD method. Prior to film deposition, the single crystal KTO substrates were annealed at 700 °C for 30 minutes under vacuum degree of $1\times10^{-8}$ Torr in PLD chamber. After the substrate treatment, the substrate temperature was maintained at 700 °C to facilitate the deposition of ETO thin films. During the thin film growth process, the energy density, frequency, and pressure of the excimer laser (KrF excimer laser with a wavelength 248 nm) were set at 1.5 J/cm$^2$, 2 Hz, and $1\times10^{-8}$ Torr respectively. The crystal structures of samples were analyzed at room temperature by an XRD (Rigaku Smart Lab) with a Ge (220)×2 crystal monochromator. The chemical valence states of the films were characterized by XPS (Thermo Scientific ESCALAB 250). The magnetic properties of the ETO/KTO heterostructures were assessed using a MPMS, while the transport properties of the 2DEGs were measured by a Physical Property Measurement System (PPMS).

Cross-sectional TEM samples were prepared by focused-ion beam equipment (Helios 600i) with Ga ions using a standard workflow. TEM characterization was performed in a double aberration-corrected JEOL Grand ARM 300F microscope. The HAADF-STEM images were acquired with collection semi-angle 54-220 mard and convergence semi-angle 22 mrad at 300 kV. For the EELS setup, a convergence semi-angle of 22 mrad, and a collection semi-angle of 76 mrad were used in this work. The dural EELS mode was taken for EELS mapping with a step of 0.55 Å, probe size of 1Å and a dwell time of 0.05 s per pixel, so that energy shift can be removed by the zero energy loss peak. The energy dispersion is 1 eV/channel to cover the big range between K-L edge and Ta-M edge. A power-law function was adapted to subtract the background of the EELS spectra and a Fourier deconvolution was utilized to remove multiple scattering. The K-L$_{2,3}$ edge, Ti-L$_{2,3}$ edge, Eu-M$_{4,5}$ edge and Ta-M$_{4,5}$ edge are used for EELS mapping images. A spot size of 8C was adopted for EELS data acquisitions with beam current of ~10 pA as we referred to the instructions of ARM 300F.

We perform DFT calculations within the plane-wave approach, as implemented in the Vienna Ab initio Simulation Package (VASP). We use projected augmented wave

(PAW) pseudopotentials with an energy cutoff of 600 eV. Eu 4*f* electrons are treated as valence electrons. To model the correlation effects on Eu-*f* orbitals, we use the rotationally invariant DFT+*U* method (by setting LDAUTYPE=2 in VASP). For bulk ETO, we use a $\sqrt{2}\times\sqrt{2}\times 2$ supercell, which is large enough to accommodate various magnetic orders. For $Eu_{0.25}K_{0.75}TaO_3$, we use a $2\times 2\times 2$ supercell. The threshold for the charge self-consistent calculations is $10^{-7}$ eV. Both cell and internal atomic positions are fully relaxed until each force component is smaller than 1 meV/Å. We use a Monkhorst-Pack **k** mesh of $12\times 12\times 12$ to sample the first Brillouin zone. We neglect the spin-orbit coupling in the calculations since it is not important for the Kondo physics and anti-weak-localization effects.

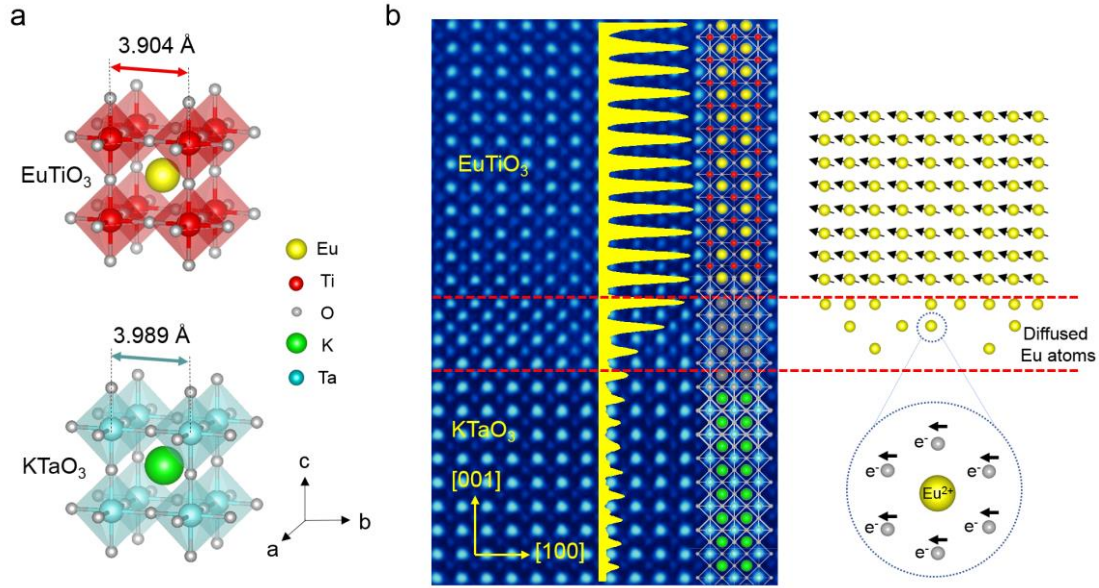

**Fig.1 Schematic diagram of the mechanism responsible for the enhancement of enhanced ferromagnetism in the 2DGE at the ETO/KTO interface.** (a) The structural diagrams of the bulk perovskite oxides, ETO and KTO, show lattice constants of 3.904 Å and 3.989 Å, respectively. (b) Left: cross-sectional HAADF-STEM image of the interface between ETO film and KTO with atomic model placed on. The atomic line profile of atoms at site A is represented by a yellow peak array. Right: schematic diagram illustrates that the integration of magnetic atom doping and the proximity effect can result in the enhanced ferromagnetism 2DEG at the ETO/KTO interface.

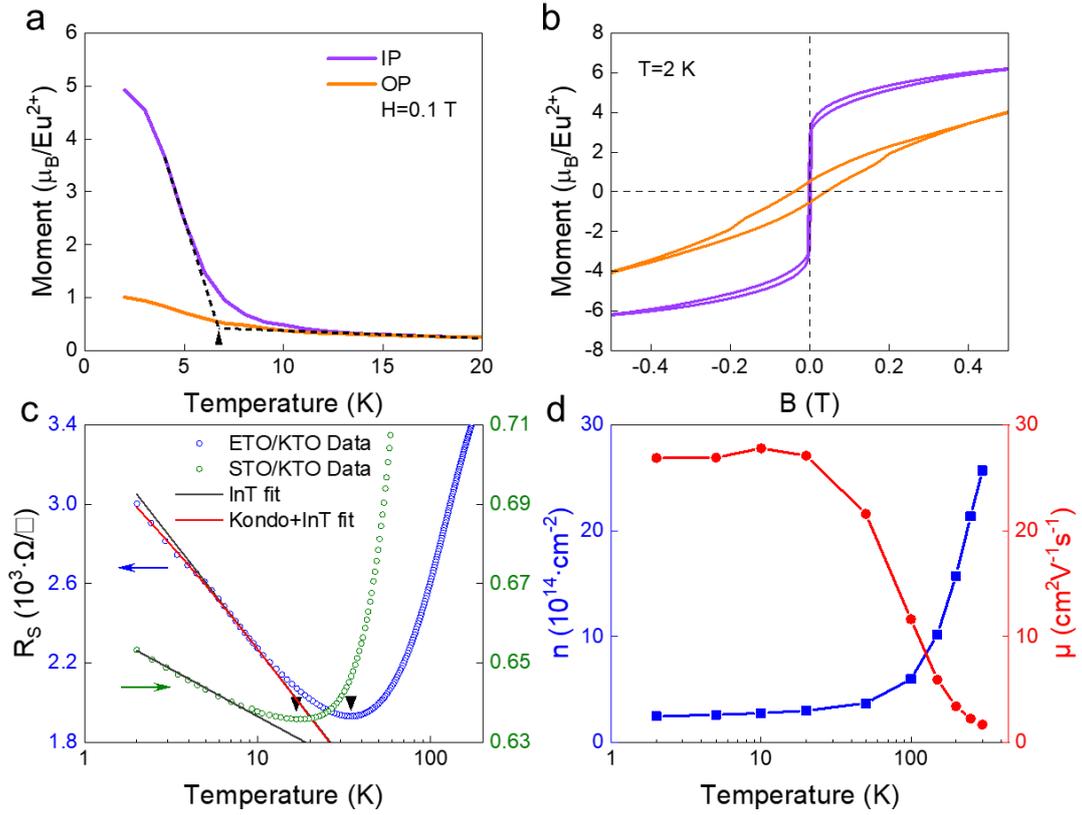

**Fig.2 Magnetic and electronic properties of the ETO/KTO heterostructure.** (a) Temperature-dependent magnetization of the ETO/KTO heterostructure measured utilizing both in-plane and out-of-plane field cooling modes. (b) Magnetic field-dependent magnetization measured at 2 K with in-plane and out-of-plane fields, respectively. The easy magnetic axis of the ETO layer lies in film plane. (c) Sheet resistance at $B = 0$ as a function of temperature on a logarithmic scale for the ETO/KTO interface. Gray line shows the low-temperature logarithmic dependences, while red line corresponds to the theoretical fits. (d) Carrier density and mobility as functions of temperature.

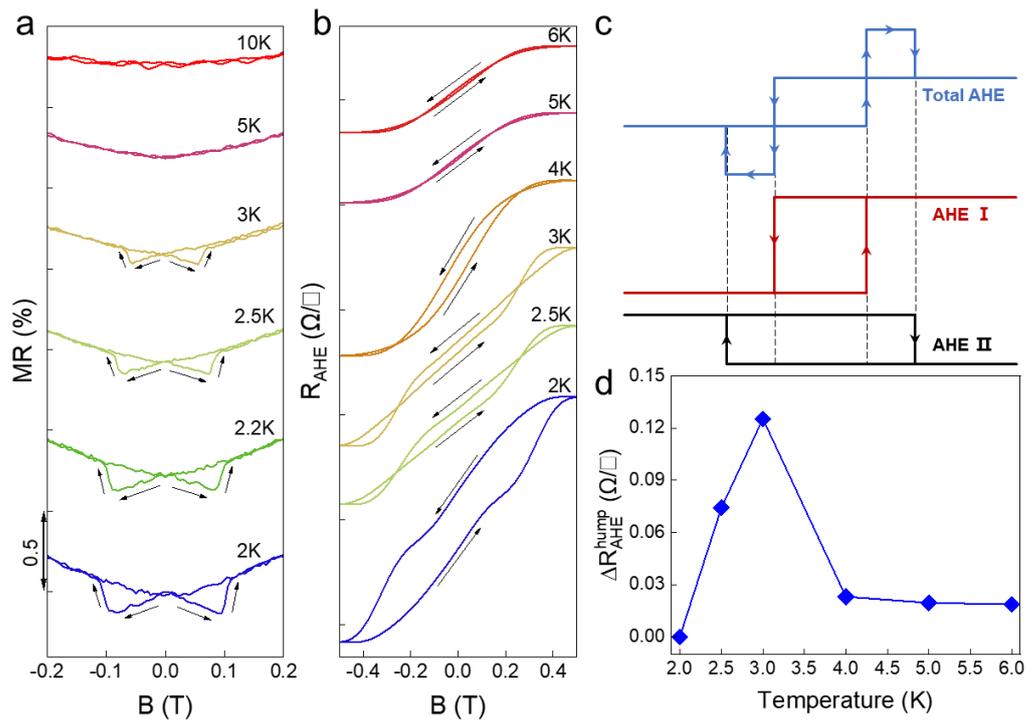

**Fig.3 Magnetotransport properties of the 2DEGs.** (a) MR of the 2DEG measured from -0.2 T to 0.2 T under varying temperatures, with notable hysteresis in magnetic resistance. (b) Anomalous Hall resistance of 2DEG as a function of magnetic field. (c) Double AHE model incorporates two AHE contributions, each with a different sign and intensity. (d) Temperature-dependent intensity of the hump in (b).

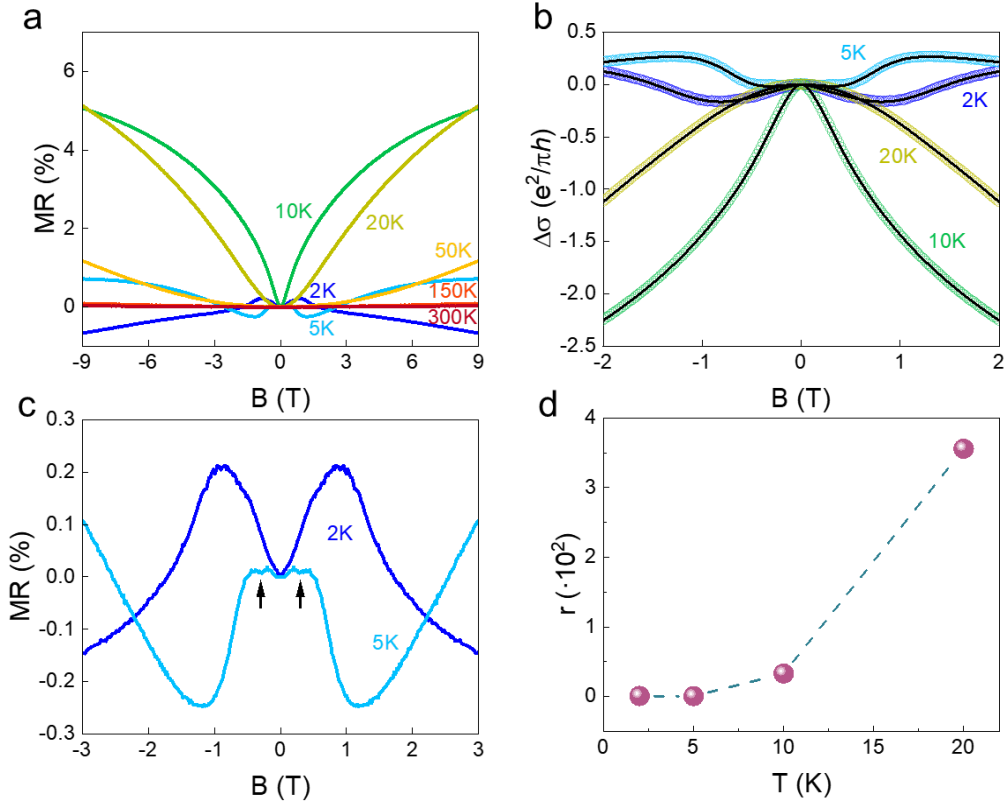

**Fig.4 The interplay between magnetic scattering and the WAL effect.** (a) Magnetoresistance of the 2DEG measured from -9 T to 9 T under varying temperatures. (b) Magnetoconductance $\triangle\sigma=\sigma(B)-\sigma(0)$, expressed in units of $e^2/\pi h$, measured from -2 T to 2 T under different temperatures. The black solid lines represent the fitting lines derived from the Eq .(3). (c) A detailed magnified examination of the magnetoresistance at temperatures of 2 K and 5 K. (d) Ratio of WAL contribution to magnetic scattering contribution in magnetoresistance as a function of temperature.

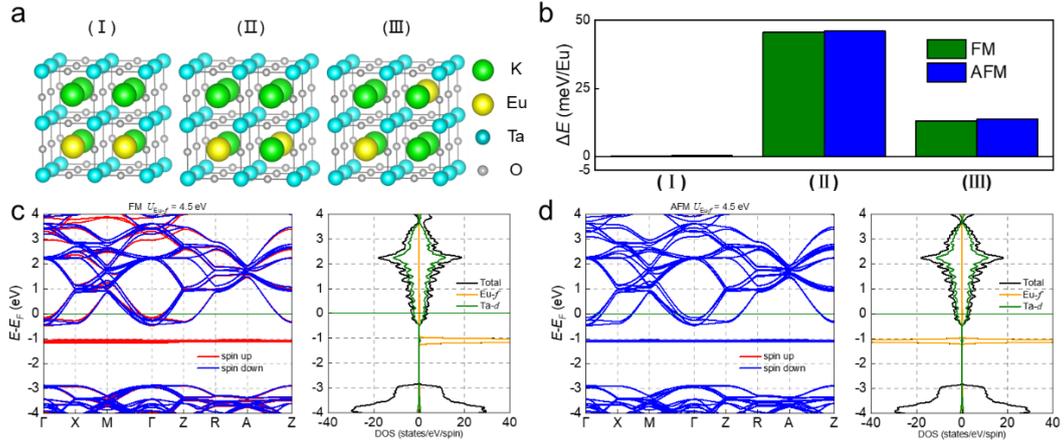

**Fig.5 The results of DFT calculations for the Eu doped KTO.** (a) The supercell of $Eu_{0.25}K_{0.75}TaO_3$, with two Eu atoms positioned in three different configurations: the two Eu atoms are (I) separated by about one lattice constant $a$; (II) along the face-diagonal line with the separation of about $\sqrt{2}a$; (III) along the body-diagonal line with the separation of about $\sqrt{3}a$. (b) Total energy of the three configurations of $Eu_{0.25}K_{0.75}TaO_3$ with Eu-$f$ moments either aligned ferromagnetically or antiferromagnetically. The calculations are done with $U=4.5$ eV on Eu-$f$ orbitals. The band structure (left) and density of states (right) for $Eu_{0.25}K_{0.75}TaO_3$ in configuration (I), calculated with $U=4.5$ eV on Eu-$f$ orbitals, for both the FM state (c) and AFM state (d).

This work was supported by the National Key Research and Development Program of China (Grant No. 2022YFA1403000), and the National Natural Science Foundation of China (Grant No. 12250710675, 12204522, 12434007 and U23A20366)